# Graphene growth on h-BN by Molecular Beam Epitaxy

Jorge M. Garcia[*,†,‡], Ulrich Wurstbauer[‡], Antonio Levy[‡], Loren N. Pfeiffer[§], Aron Pinczuk[‡,**], Annette S. Plaut[††], Lei Wang[‡‡], Cory R. Dean[§§], Roberto Buizza[‡‡], Arend M. Van Der Zande[‡‡], James Hone[‡‡], Kenji Watanabe[***], and Takashi Taniguchi[***]

ABSTRACT

The growth of single layer graphene nanometer size domains by solid carbon source molecular beam epitaxy on hexagonal boron nitride (h-BN) flakes is demonstrated. Formation of single-layer graphene is clearly apparent in Raman spectra which display sharp optical phonon bands. Atomic-force microscope images and Raman maps reveal that the graphene grown depends on the surface morphology of the h-BN substrates. The growth is governed by the high mobility of the carbon atoms on the h-BN surface, in a manner that is consistent with van der Waals epitaxy. The successful growth of graphene layers depends on the substrate temperature, but is independent of the incident flux of carbon atoms.

[*] Corresponding author: jm.garcia@csic.es
[†] MBE Lab, IMM-Instituto de Microelectrónica de Madrid (CNM-CSIC),Isaac Newton 8, PTM, E-28760 Tres Cantos, Madrid, Spain
[‡] Department of Physics, Columbia University, New York, NY, USA
[§] Electrical Engineering Department, Princeton University, NJ, USA
[**] Department of Applied Physics and Applied Mathematics, Columbia University, New York, NY, USA
[††] School of Physics, Exeter University, Exeter, UK
[‡‡] Department of Mechanical Engineering, Columbia University, NY, NY, USA
[§§] Department of Electrical Engineering and Department of Mechanical Engineering, Columbia University, NY, NY, USA
[***] Advanced Materials Laboratory, National Institute for Materials, Science, 1-1 Namiki, Tsukuba, 305-0044, Japan.

Studies of atomic layers of graphene attract enormous interest for their impact in fundamental science and for their potential to revolutionize applications in diverse areas such as electronics and optoelectronics [1, 2]. Much of the exciting research has been reported on high quality graphene obtained by micromechanical exfoliation of graphite. Advances in fundamental and applied research and technology would be greatly enhanced by implementation of scalable fabrication methods. While chemical vapor deposition (CVD) creates the potential for production of large area graphene layers [3, 4], device performance of CVD grown films remains low with mobilities as much as 10 times smaller than that measured in exfoliated devices. A promising alternative is the growth of graphene by molecular beam epitaxy (MBE). MBE growth of graphene on substrates patterned at the nanoscale, to give an example, could lead to fabrication of nanostructures with controlled doping and energy gaps [5, 6, 7, 8]. The remarkable potential of MBE grown graphene has resulted in numerous challenging developments [9, 10, 11, 12, 13].

Single crystal hexagonal boron nitride (h-BN) has been proposed as an ideal substrate for epitaxial growth of graphene by MBE [14, 15]. The hexagonal lattice structure yields an atomically flat surface with lattice constant close to that of graphene (less than 2% mismatch). h-BN is also an inert large-band-gap insulator that can withstand very high temperatures. Further, it has been recently demonstrated that h-BN is an ideal substrate for electrical transport devices fabricated from exfoliated graphene flakes [16, 17, 18].

In this letter we report the MBE growth of single layer graphene on single crystal h-BN flakes. Characterization of the MBE grown graphene layers by Raman-scattering spectroscopy and atomic-force microscope (AFM) imaging indicate that the graphene layers consist of nanoscale domains. The non-uniformity of the growth suggests that the individual characteristics of the h-BN flakes, such as surface morphology, may play a significant role. The maximum substrate temperature that can be reached before the $SiO_2$ substrate decomposes is a key parameter limiting the growth conditions in this work. The quality of the layers depends critically on the substrate temperature during growth. The best results are obtained at growth temperature of $T_G$=930°C, where Raman characterization reveals optical phonon spectra that are very similar to graphene nanoribbons (GNRs) fabricated from etched exfoliated pristine graphene [19, 20]. These results demonstrate the great potential of graphene MBE in studies of the remarkable physics and materials science of these atomic layers.

We find that the growth is independent of the flux of carbon atoms. This is a characteristic feature of graphene MBE growth on h-BN which is attributed to the high mobility of the carbon atoms on the h-BN surface. This is a key property which ultimately dictates the characteristics of carbon MBE growth on h-BN, and which determines the character of graphene nanodomains on h-BN. This finding of high carbon atom mobility is believed to be a consequence of the weak van der Waals interaction between the carbon atoms and the h-BN substrate surface. Our results thus indicate that MBE growth of graphene on h-BN is a form of van der Waals epitaxy [21].

The graphene layers are grown in a MBE ultra-high vacuum chamber of custom design and construction that is equipped with a solid carbon source in close proximity to the substrate [13, 22, 23]. The inset to Fig.1 schematically describes the relative positions of the solid carbon source and the substrates. Figure 1 shows the normalized flux $\Phi_N(d)=\Phi(d)/\Phi(0)$ as a function of the position $d$ on the substrate (see inset to Fig. 1). The calibration of $\Phi(d)$ is obtained by measuring, with an AFM, the position-dependent thickness of a thick carbon film grown on $SiO_2$. The red line in Fig. 1 is a fit to a cosine square law [13, 22, 23]. This geometry provides a large gradient in the carbon flux $\Phi_N(d)$.

H-BN flakes, obtained by mechanical exfoliation, are deposited on a 300nm thick layer of $SiO_2$ on Si. The h-BN flakes have a large variation in lateral size (from 10 to 150 µm), thickness and shape. The flakes examined in this study show a dark-blue color when visualized in an optical microscope under white light illumination. AFM measurements indicate that thickness of these h-BN flakes is 10-15 nm. Prior to growth, the samples are degassed *in situ* for 5 minutes at 380°C. The growth time is 40.6 minutes, corresponding to 5 nm of carbon at *d*=0. The substrate temperature during growth, $T_G$, can be adjusted with a heater on the back of the sample. $T_G$ was varied between 600°C and 930°C, with the upper value of $T_G$ limited by the decomposition temperature of $SiO_2$.

Raman spectra are obtained with a Renishaw *in-via* Micro-Raman instrument, that is equipped with a movable x-y-z stage and a 532 nm laser for excitation of the spectra. The laser is focused to a spot of 0.5 µm diameter at a power of less than 3 mW. The spectra display the optical phonon Raman bands of graphene layers and h-BN. They overlap a broad photoluminescence background, which appears after *in situ* degassing the h-BN flakes. In all the Raman measurements shown here, the background is subtracted by means of a polynomical function, and the spectra are normalized to the integrated intensity of the silicon phonon bands.

Raman characterization of the graphene MBE growth are acquired as a function of carbon flux by taking advantage of the growth geometry shown in Fig. 1. Surprisingly, we find no significant dependence of the peaks of the Raman spectra on $\Phi_N(d)$. This is a striking result that can be understood as a manifestation of the high mobility of the carbon atoms on the h-BN surface. We present below AFM images that are consistent with this interpretation.

Figures 2 (a) and 2(b) show Raman spectra from MBE growth at $T_G$=930°C. Figure 2(a) is from the region labeled A in the inset to Fig. 2(a). The spectrum in Fig. 2(b) is from the region labeled B in the same inset to Fig. 2(a). The much broader Raman bands in the spectrum shown in Fig. 2(b) and the absence of Raman bands characteristic of single layer graphene indicate that carbon MBE deposition here yields much lower quality graphitic layers. [24, 25]. Samples grown at lower temperatures ($T_G$=600°C, 750°C, and 850°C) show Raman spectra that are similar to those of area B, which do not show a clear signature of single layer graphene.

The spectrum in Fig. 2(a), arising from region A, has a doublet at about 1350 $cm^{-1}$ which is the superposition of the Raman D band of graphene (defect activated) and the sharper TO phonon band of h-BN [26]. The doublet at about 1600 $cm^{-1}$ is from the superposition of the G and D' modes of single layer graphene [27, 28]. The G band is a long wavelength optical phonon and the D' is the critical point in the optical phonon dispersion which is due to relaxation of wave-vector conservation rules induced by defects [27, 28]. The 2D band at about 2680 $cm^{-1}$ (a.k.a. G') is from second-order Raman scattering by two optical phonons [27, 28]. At even higher energy shifts the spectrum reveals three distinct second-order Raman bands [27, 29]. All the bands shown in Fig. 2(a) have been fitted with Lorentzian line shape functions. Their frequencies and FWHMs are listed in Table 1, where these values are compared with those reported for 30 nm GNRs [20].

While the presence of intense Raman D bands is normally linked to defects, the D band in Fig. 2(a) exhibits a striking similarity to those reported in GNRs obtained by patterning and etching mechanically exfoliated single layer graphene. In GNRs the large intensities and characteristic sharpness of the D Raman band is attributed to an increase in the fraction of edge carbons that act as defects [19, 20]. The similarity of the Raman D bands in MBE graphene and GNRs thus indicates that the graphene layers in MBE growth consist of domains with nanoscale dimensions exhibiting a large fraction of edge carbons. Table I shows quantitatively the similarities between the Raman D bands in the MBE graphene and that of GNRs of 30 nm width. The large integrated intensity ratio of the D and G bands, I(D)/I(G) = 3.6, and a narrow D band (FWHM=29 cm$^{-1}$) are characteristic of narrow GNRs of widths in the range 30 – 50 nm [19, 20]. This comparison suggests that the growth in region A consists of nanoscale domains of single-layer graphene

The number of graphene layers can be determined from the lineshape analysis of the second-order 2D band. The 2D peak shown in Fig. 2(a) is well fit by a single Lorentzian with a FWHM of 53cm$^{-1}$, which is comparable to the FWHM of the 2D band measured in single layer GNRs [19, 20]. In Fig. 2(a) the integrated intensity peak ratio I(2D)/I(G) is approximately two. A ratio I(2D)/I(G) larger than one is commonly interpreted as a signature of single layer graphene [27, 28, 30]. While exchange of charge between the grown graphene layers with the h-BN substrate may strongly influence the I(2D)/I(G) ratio [31], the marked similarities in Raman spectra with those from single layer GNR are regarded as evidence of the growth of single-layer graphene nanodomains.

Figures 3(a) – 3(c) are micro Raman maps of the flake in the inset to Fig. 2. Figures 3(d) and (e) are AFM images. Figure 3(a) shows the spatial variation of I(2D). The bright area in the map reveals region A as a micrometer-scale area characterized by a I(2D)/I(G) ratio that is larger than 2, as shown in Fig. 3(b). Figure 3(c) reveals a I(D)/I(G) ratio of about 4, as mentioned above in the discussion of the results of Fig. 2.

The black square in Fig. 3 (d) highlights a segment of region A of area 1.5x1.5 $\mu m^2$. This area is presented at much greater magnification in Fig. 3(e). The image in Fig. 3(e) reveals a highly non-uniform growth that consists of nanoscale domains. Height profiles constructed from these AFM data reveal that some of the domains have a layer thickness that is expected for single-layer graphene (about 0.33 nm). We surmise that these are the single-layer graphene domains that contribute to Raman spectra such as those in Fig. 2(a).

In the AFM images it is clearly visible that some areas of the h-BN flake do not show deposited carbon. These depleted regions are the ones with absence of graphene Raman signatures in Figs. 3(a) – 3(c). Further, the image in Fig. 3(d) reveals that large amounts of carbon accumulate on some of the flat terraces and particularly on the edges of the flake. These findings are interpreted as a manifestation of the high mobility of the carbon atoms on the h-BN surface. This large carbon mobility, which manifests as the lack of dependence of the Raman peaks on $\Phi_N(d)$, results in migration and nucleation of the carbon at surface steps and edges.

In summary, we have demonstrated the MBE growth of graphene nanometer size domains on h-BN single crystal flakes - acting as substrates. The Raman signatures from these single-layer graphene domains are similar to those from graphene nanoribbons fabricated from mechanically exfoliated graphene. AFM measurements show monolayer graphene steps and non-uniform coverage of the grown graphene. We observe negligible dependence of the Raman peaks on carbon flux which we attribute to the high mobility of the carbon atoms on the h-BN. These results open up the possibility of controlled growth of graphene by MBE on h-BN.

**Acknowledgments**

This work is supported by ONR (N000140610138 and Graphene MURI), AFOSR (FA9550-11-1-0010), EFRC Center for Re-Defining Photovoltaic Efficiency through Molecule Scale Control (award DE-SC0001085), NSF (CHE-0641523), NYSTAR, CSIC-PIF (200950I154), Spanish CAM (Q&C Light (S2009ESP-1503), Numancia 2 (S2009/ENE-1477)) and Spanish MEC (ENE2009-14481-C02-02, TEC2011-29120-C05-04, MAT2011-26534).

Figure captions:

Figure 1: Normalized flux of carbon atoms as function of position on the substrate ($d$). The inset shows a schematic side view of the MBE geometry ($D_0$=15 mm). The growth temperature $T_G$ is measured with a thermocouple (tc) at $D_0$. The flux gradient along $d$ (see equation) is plotted as the red line.

Figure 2: Raman analysis of MBE graphene on a h-BN flake, grown at $T_G$=930°C. (a) Spectrum from position A showing the distinctive Raman signatures of single-layer graphene . The inset shows an optical microscope image of the flake. (b) Spectrum from position B, interpreted as coming from low quality graphitic layers.

Figure 3: Raman and AFM images of the flake in the inset of Fig. 2(a). The spatial resolution of the Raman maps is 0.5 µm. (a) Map of the integrated intensity of the 2D band. (b) Map of the integrated intensity ratio of the D and G bands. (c) Map of the integrated intensity ratio of the 2D and G bands. (d) 20x15.6 µm$^2$ AFM image. (e) 1.5x1.5 µm$^2$ AFM image of the region A area showing the single-layer graphene nanodomains. Height profiles reveal that the bright elongated areas are graphene wrinkles.

Table 1: Fitting parameters of the Raman spectra. The left panel is for Fig. 2(a). The right panel is for 30 nm GNR (after Ref. [20]).

Table 1

|     | Graphene on h-BN: Region A | | | 30nm GNR (after Ref. [20]) | | |
| --- | --- | --- | --- | --- | --- | --- |
|     | $\omega$ (cm$^{-1}$) | FWHM (cm$^{-1}$) | I(peak)/I(G) | $\omega$ (cm$^{-1}$) | FWHM (cm$^{-1}$) | I(peak)/I(G) |
| D   | 1345 | 29 | 3.6 | 1340 | ~28 | 6 |
| G   | 1589 | 42 | -   | ~1586 |    |   |
| D'  | 1622 | 21 | -   | 1620 |    |   |
| 2D  | 2680 | 53 | 2.0 | 2680 | 50 |   |

# References

[1] A. K. Geim, K.S. Novoselov, *The rise of graphene*. Nature Materials, **6** (2007) 183-191.

[2] M.H. Rümmeli, C. G. Rocha, F. Ortmann, I. Ibrahim, H. Sevincli, F. Börrnert, J. Kunstmann, A. Bachmatiuk, M. Pötschke, M. Shiraishi, M. Meyyappan, B. Büchner, S. Roche, G. Cuniberti, *Graphene: Piecing it Together. Advanced Materials*, **23** (2011) 4471-4490.

[3] X. Li, W. Cai, J. An, S. Kim, J. Nah, D. Yang, R. Piner, A. Velamakanni, I. Jung, E. Tutuc, S. K. Banerjee, L. Colombo, R. S. Ruoff, *Large-Area Synthesis of High-Quality and Uniform Graphene Films on Copper Foils.* Science, **324** (2009) 1312-1314.

[4] K.S. Kim, Y. Zhao, H. Jang, S. Y. Lee, J. M. Kim, K. S. Kim, J.-H Ahn, P. Kim, J.-Y Choi, B. H. Hong, *Large-scale pattern growth of graphene films for stretchable transparent electrodes*. Nature, **457** (2009) 706-710.

[5] Y.-W. Son, M. L. Cohen, and S. G. Louie, *Energy Gaps in Graphene Nanoribbons*, Phys. Rev. Lett. **97** (2006) 216803.

[6] Melinda Y. Han, Barbaros Ozyilmaz, Yuanbo Zhang, and Philip Kim, *Energy Band-Gap Engineering of Graphene Nanoribbons*, Phys. Rev. Lett. **98** (2007) 206805.

[7] X. Wang, Y. Ouyang, X. Li, H.Wang, J. Guo, H. Dai, *Room-Temperature All-Semiconducting Sub-10-nm Graphene Nanoribbon Field-Effect Transistors*. Phys. Rev. Lett. **100** (2008) 20.

[8] C. Stampfer, J. Güttinger, S. Hellmüller, F. Molitor, K. Ensslin, and T. Ihn, *Energy Gaps in Etched Graphene Nanoribbons*, Phys. Rev. Lett. **102** (2009) 056403.

[9] J. Park, W. C. Mitchel, L. Grazulis, H. E. Smith, K. G. Eyink, J. J. Boeckl, D. H. Tomich, S. D. Pacley, J. E. Hoelscher, *Epitaxial Graphene Growth by Carbon Molecular Beam Epitaxy (CMBE)*. Adv. Mater., **22** (2010) 4140-4145.

[10] E. Moreau, S. Godey, F. J. Ferrer, D. Vignaud, X. Wallart, J. Avila, M. C. Asensio, F. Bournel, J. J. Gallet, *Graphene growth by molecular beam epitaxy on the carbon-face of SiC*, Appl. Phys. Lett. **97** (2010) 241907.

[11] S. K. Jerng, D. S. Yu, Y. S. Kim, J. Ryou, S. Hong, C. Kim, S. Yoon, D. K. Efetov, P. Kim, S. H. Chun, *Nanocrystalline Graphite Growth on Sapphire by Carbon Molecular Beam Epitaxy.* J. Phys. Chem. C, **115** (2011) 4491-4494.

[12] G. Lippert, J. Dabrowski, M. Lemme, C. Marcus, O. Seifarth, G. Lupina, *Direct graphene growth on insulator*. Phys Status Solidi (b) **248** (2011) 2619-2622.

[13] U. Wurstbauer et al., arXiv:1202.2905v1 [cond-mat.mtrl-sci], (2012).

[14] J.M Garcia, L. N. Pfeiffer, *Devices with graphene layers*. International App. No.: PCT/US2008/013796. (2009)

[15] L. N. Pfeiffer, *Devices including graphene layers epitaxially grown on single crystal substrates*. United States Patent number 7619257 (2009)

[16] J. Xue, J. Sanchez-Yamagishi, D. Bulmash, P. Jacquod, A. Deshpande, K. Watanabe, T. Taniguchi, P. Jarillo-Herrero, B. J. LeRoy, *Scanning tunnelling microscopy and spectroscopy of ultra-flat graphene on hexagonal boron nitride*. Nat Mater, **10** (2011) 282-285.

[17] C. R. Dean, A. F. Young, I. Meric, C. Lee, L. Wang, S. Sorgenfrei, K. Watanabe, T. Taniguchi, P. Kim, K. L. Shepard, J. Hone, *Boron nitride substrates for high-quality graphene electronics*, Nat Nano, **5** (2010) 722-726.

[18] Liam Britnell, Roman V. Gorbachev, Rashid Jalil, Branson D. Belle, Fred Schedin, Mikhail I. Katsnelson, Laurence Eaves, Sergey V. Morozov, Alexander S. Mayorov, Nuno M. R. Peres, Antonio H. Castro Neto, Jon Leist, Andre K. Geim, Leonid A. Ponomarenko, Kostya S. Novoselov, *Atomically thin boron nitride: a tunnelling barrier for graphene devices*, arXiv:1202.0735v1 (2012).

[19] Sunmin Ryu, Janina Maultzsch, Melinda Y. Han, Philip Kim, and Louis E. Brus, *Raman Spectroscopy of Lithographically Patterned Graphene Nanoribbons*, ACS Nano, **5** (2011) 4123–4130.

[20] D. Bischoff, J. Güttinger, S. Dröscher, T. Ihn, K. Ensslin, and C. Stampfer, *Raman spectroscopy on etched graphene nanoribbons*, J. Appl. Phys. **109** (2011) 073710.

[21] A. Koma, J. Cryst. Growth **201-202**, 236 (1999); A. Koma, *Van der Waals epitaxy—a new epitaxial growth method for a highly lattice-mismatched system*, Thin Solid Films, **216** (1992) 72-76.

[22] Jorge M. Garcia, Rui He, Mason P. Jiang, Jun Yan, Aron Pinczuk, Yuri M. Zuev, Keun Soo Kim, Philip Kim, Kirk Baldwin, Ken W. West, Loren N. Pfeiffer, *Multilayer graphene films grown by molecular beam deposition.* Solid State Commun., **150** (2010) 17–18.

[23] Jorge M. Garcia, Rui He, Mason P. Jiang, Philip Kim, Loren N. Pfeiffer, Aron Pinczuk; *Multilayer graphene grown by precipitation upon cooling of nickel on diamond*, Carbon **49** (2011) 1006–1012.

[24] A. C. Ferrari, J. Robertson, *Interpretation of Raman spectra of disordered and amorphous carbon*. Phys. Rev. B, **61** (2000) 14095.

[25] A. C. Ferrari, J. Robertson, *Raman spectroscopy of amorphous, nanostructured, diamond–like carbon, and nanodiamond*. Phil. Trans. of the Royal Society of London. Series A; Mathematical, Physical and Engineering Sciences, **362** (2004) 2477-2512.

[26] S. Reich, A. C. Ferrari, R. Arenal, A. Loiseau, J. Robertson, *Resonant Raman scattering in cubic and hexagonal boron nitride,* Phys. Rev. B **71** (2005) 205201

[27] M. A. Pimenta, G. Dresselhaus, M. S. Dresselhaus, L. G. Cançado, A. Jorio and R. Saito, *Studying disorder in graphite-based systems by Raman spectroscopy*, Phys. Chem. Chem. Phys., **9** (2007) 1276

[28] Andrea C. Ferrari, *Raman spectroscopy of graphene and graphite: Disorder, electron–phonon coupling, doping and nonadiabatic effects*, Solid State Commun. **143** (2007) 47–57.

[29] E. H. Martins Ferreira, Marcus V. O. Moutinho, F. Stavale, M. M. Lucchese, Rodrigo B. Capaz, C. A. Achete, and A. Jorio, *Evolution of the Raman spectra from single-, few-, and many-layer graphene with increasing disorder*, Phys. Rev. B 82 (2010) 125429.

[30] D. Graf, F. Molitor, K. Ensslin, C. Stampfer, A. Jungen, C. Hierold, and L. Wirtz, *Spatially Resolved Raman Spectroscopy of Single- and Few-Layer Graphene*, Nano Lett., **7**(2007) 238.

[31] A. Das, S. Pisana, B. Chakraborty, S. Piscanec, S. K. Saha, U. V. Waghmare, K. S. Novoselov, H. R. Krishnamurthy, A. K. Geim, A. C. Ferrari, and A. K. Sood, *Monitoring dopants by Raman scattering in an electrochemically top-gated graphene transistor*, Nature Nanotechnology **3** (2008) 210 – 215.

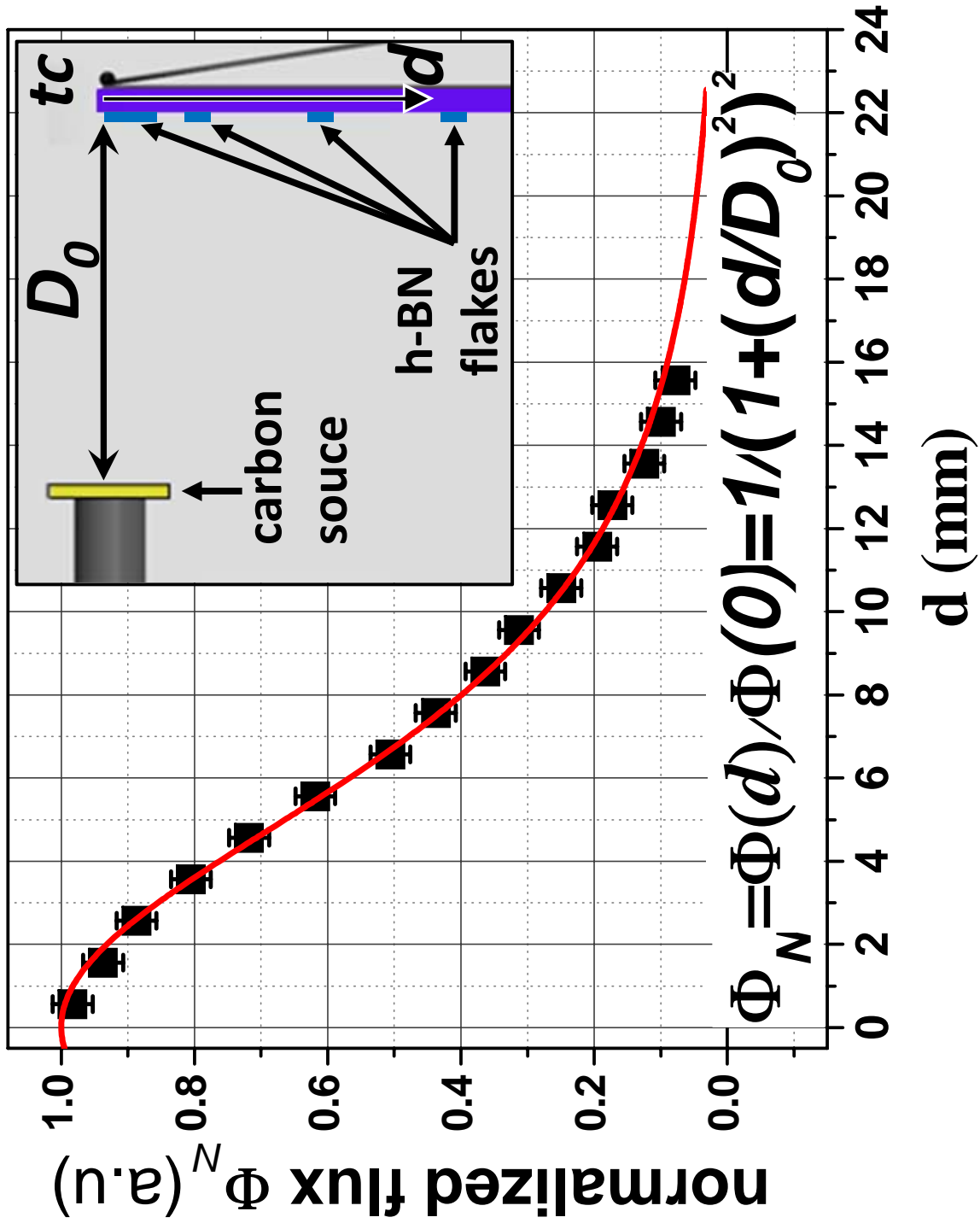

Figure 1

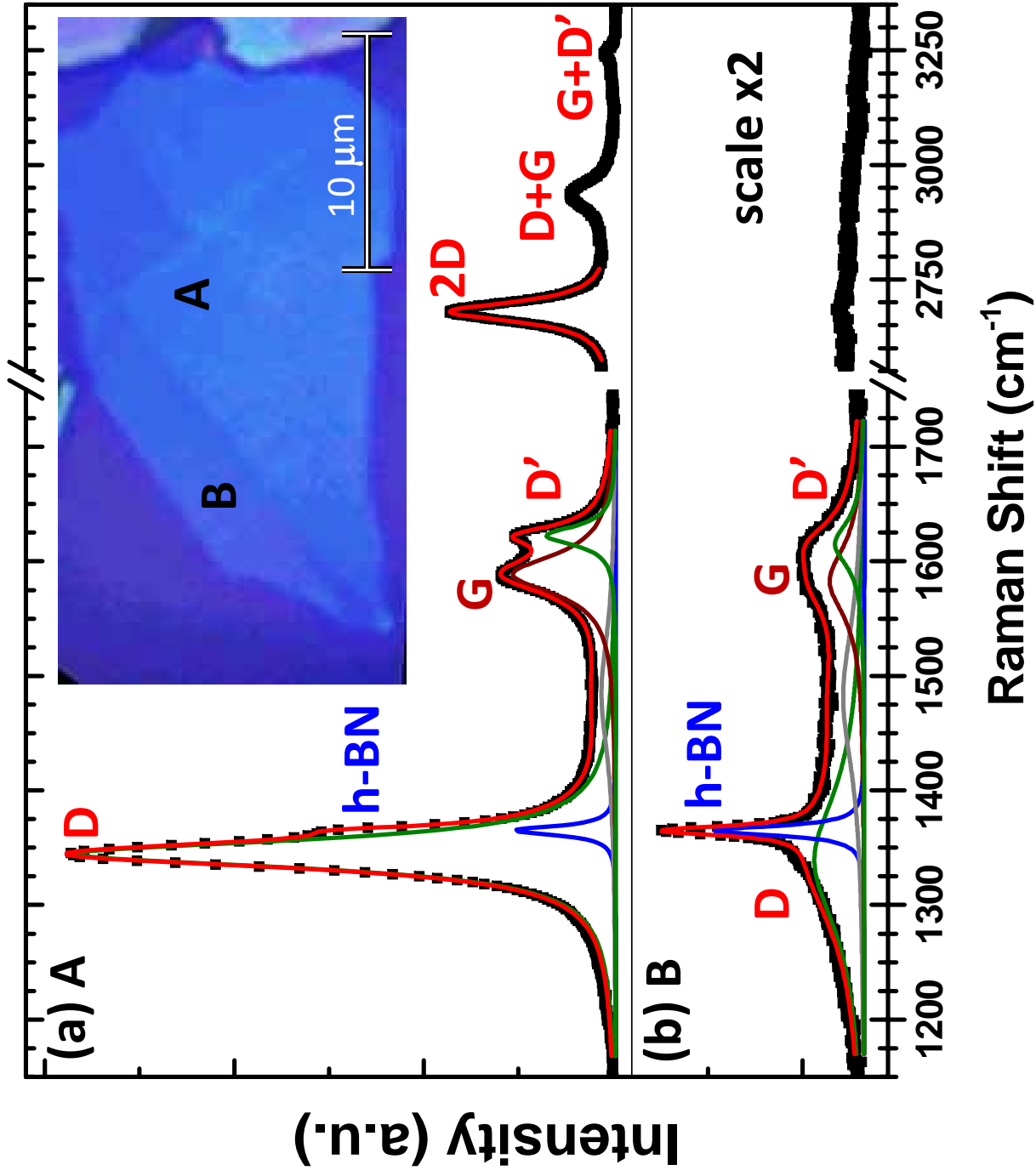

Figure 2

Figure 3

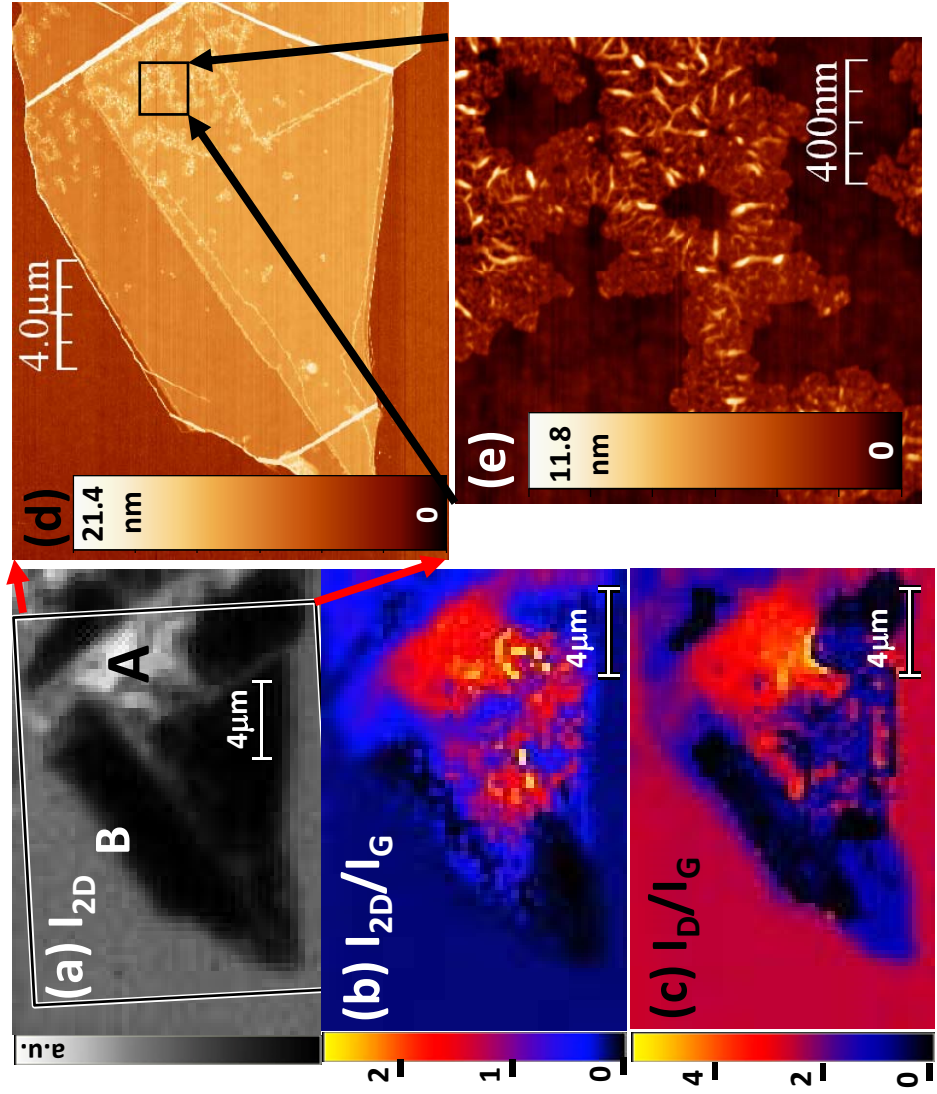